# Left ventricular myocardium of heart as a 3D scheme

Saeed Ranjbar [1] Ph.D, Mersedeh Karvandi [2,*] MD

**Research institute:**
1- Institute of Cardiovascular Research, Modarres Hospital, Shahid Beheshti University of Medical Sciences, Tehran, IR Iran
2- Taleghani Hospital, Shahid Beheshti University of Medical Sciences,Tehran, Iran

**\*Corresponding author:**
Mersedeh Karvandi
Taleghani Hospital, Shahid Beheshti University of Medical Sciences,Tehran, Iran
E-mail: mersedeh_karvandi@ipm.ir
Tel.: 00982123031331
Fax: 00982122432576

# Left ventricular myocardium of heart as a 3D scheme

**Introduction:**

In this paper it will be shown that according to the scheme theory in algebraic geometry, human heart as an elastic body can be represented as a 3D scheme, on account of algebraic equations of the myocardial fibers as the local parts of the global scheme. It is possible that the fiber movements to be discussed here are identical with the so-called "myocardial fiber transactions"; however the information available to me regarding the latter is lacking in precision, that I can from no judgment in the matter. It is hoped that some enquirer may succeed shortly in this introduced scheme suggested here, which is so important in connection with the theory of schemes.

**Method and Result:**

I draw a general algorithm of this idea and I intend to give a brief discussion for each part in the mentioned algorithm step by step.

1. A regional left ventricular wall motion study based on non-linear dimensionality reduction methods (utilizing NLDR software).
2. A fibered modeling validation of the left ventricle using reconstructed curves at part 1 ( global data)
3. Body forces "F" of landmark and contact points on the obtained fibers using of the mathematical elasticity theory.

### 1- A regional left ventricular wall motion study based on non-linear dimensionality reduction method (utilizing NLDR software):

Images were acquired from the end of diastole to the end of systole at different phases:
1- Isovolumic relaxation time, 2- Rapid filling, 3- Diastasis, 4- Atrial contraction, 5- Isovolumic contraction , 6- Ejection time and 7- the end of systole at 4 apical chamber (4C) and short-axis (SX) views within a cardiac cycle by an echocardiography machine. Obtaining worthy images at denoted phases will depend on the operator (for the best clinical examination). Short-axis views are used to evaluate the radial, circumferential, and rotatory data of the left ventricular myocardial sample. The regions of left ventricular myocardium are divided to 13 points and 12 points at 4C and SX views respectively Figure 1.
Echocardiography images play the role of observed data. We attach a sequence of images to each LV myocardial divided regions respectively and they are symbolized by $x_{i,r}$'s, and there is a sequence of observed data $x_{1,r}, x_{2,r}, \dots, x_{N,r}$, where N is the number of obtained images and r is one of the numbered regions. In fact, we have made a chain of displacements, rotations, and pure strain or non-rigid transformations (deformations) of these observed data that is started at $x_{1,r}$ and is terminated at $x_{N,r}$ ($x_{1,r} \to x_{2,r} \dots \to x_{N-1,r} \to x_{N,r}$). These chains have some conceptual interpretations of the regional LV fiber arrangement/movement (Tenenbaum et al., 2000). Acquired images $x_{1,r}, x_{2,r}, \dots, x_{N,r}$ are embedded to a meshed surface sized number of pixels of an image. Translations, rotations, and pure deformations of $x_{1,r}, x_{2,r}, \ldots, x_{N,r}$ have been implemented and studied at this surface utilizing the mathematical elasticity theory (Sokolnikoff, 1983). A distance metric is defined to compute distances between observed points. Medical interpretations of observed points over the time should be checked/and stated clearly. Images are not usual images they are images from the left ventricle. Motion (displacement and velocity), deformation (strain and strain rate), and torsion of each myocardial sample have to be extracted during a cardiac cycle in the mentioned surface. These are used on the creation of a graph G with observed data as its vertices (Saxena et al., 2004; Ranjbar and Karvandi, 2014).

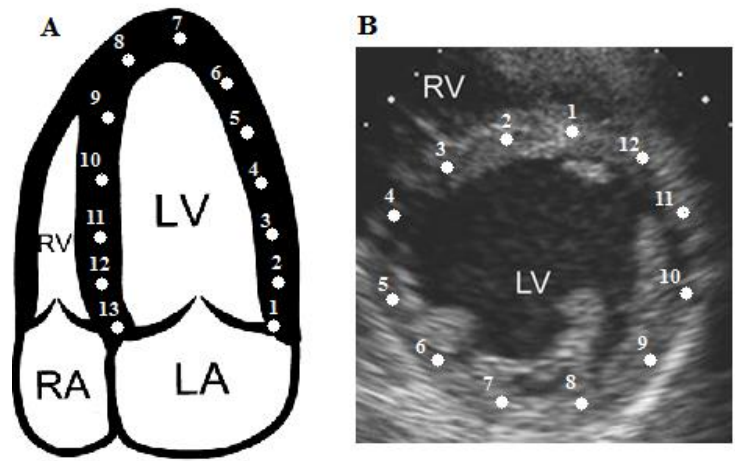

**Figure 1:** A) LV myocardial region has been divided to 13 material points at a 4C view. B) LV myocardial region has been divided to 12 material points at a SX view.

The main tool in the NLDR methods is the function "$f$" which is an isometric/conformal map. By roughly speaking the embedding "$f$" is optimized to preserve the local configurations of consecutive data sets (De Silva and Tenenbaum, 2002). A curve is reconstructed in a 3D manifold space crossing new data $y_{i,r}$'s ; $f(x_{i,r}) = y_{i,r}$ Figure 2.

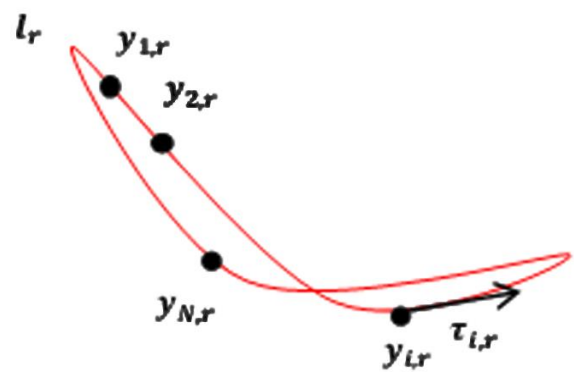

**Figure 2:** A reconstructed curve $l_r$ on a 3D manifold space and hidden data $y_{i,r}$'s ; $f(x_{i,r}) = y_{i,r}$. r is a numbered sample region and $\tau_{i,r}$ is the tangent unit vector in point $y_{i,r}$.

## 2- A fibered modeling validation of the left ventricle ( global data):

Gluing together of reconstructive curves $l_r$'s for each LV divided myocardial regions r, it gives a whole fibered structure of the left ventricle.

## 3- Body force vectors of landmark and contact points on the obtained fibers using of the mathematical elasticity theory:

The physical properties of tissue materials are closely linked to their microstructure. In order to characterize their microstructures as effectively as possible, all fibers in a sample should be individualized. This work mainly aims at developing a new fiber analysis method that segments a 3D mathematical left ventricular modeling based on echocardiography image into a background and a set of connected components, each representing a single fiber. Properly completed fiber analyses may provide input data for generative or synthetic models, which can in turn be used to estimate various characteristics of the material. In this paper, we introduce an original method based on the skeletonization of the fiber mass, followed by a geometrical analysis of the obtained skeleton. Based on this procedure, several measurements can be computed (e.g. length, translation, orientation, pure deformation, number of contacts, body force). Myocardial point connections together can be realized by a geometrical point of views like a polygon Figure 3.

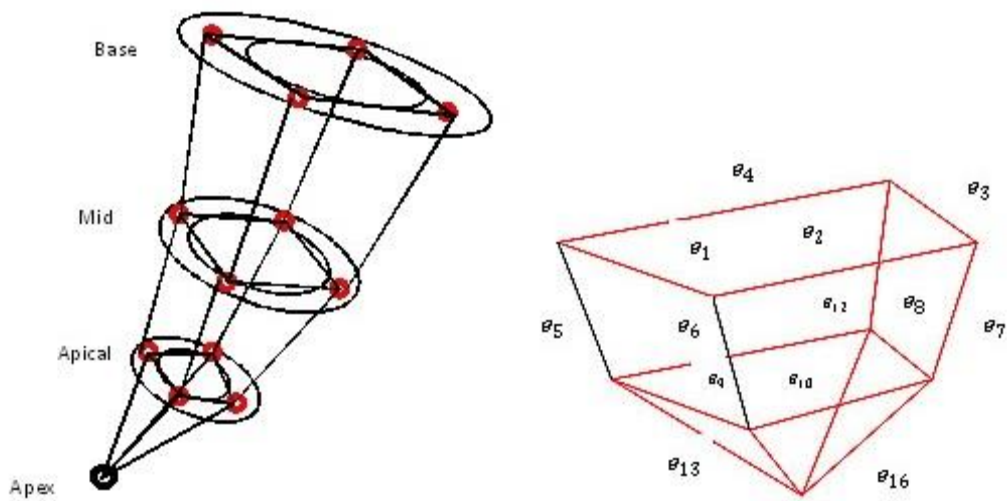

**Figure 3:** Red points are corresponding geometrical points to material points of LV myocardium. Material points are usually selected as contact points in the myocardium.

Figure 3 depicts a simplicial complex which their vertices are geometrical points corresponding to the material points and edges show how the myocardial points are connected together.

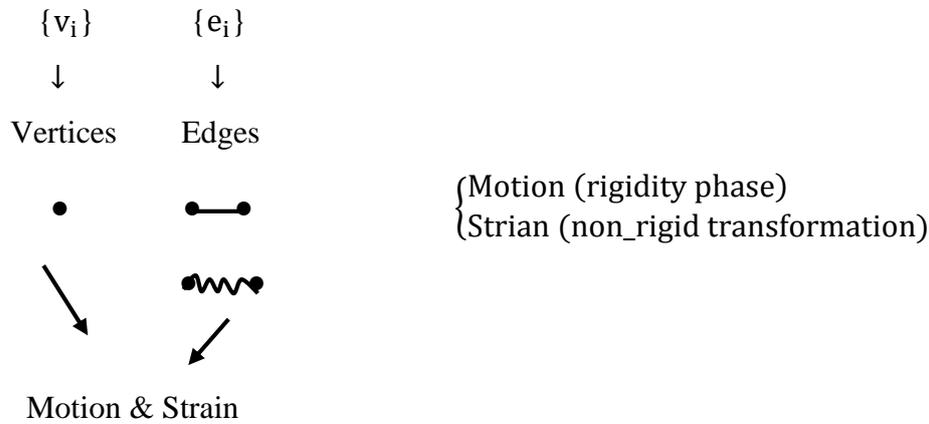

Utilizing reconstructed curves referred to 1, we recover their algebraic equations by the following:

Let p be an arbitrary point on a reconstructed curve $l_{p,r}$ which is presented with respect to the Cartesian coordinates $(x_{1,r}, x_{2,r}, x_{3,r})$ and $\tau_{p,r}$ is the unit tangent vector on it. Body force magnitude $F_r$ at p is reformulated based on pure strain components $(e_{ij,\ell_{p,r}})$ and non-rigid transformations $\tau_{p,r}$ and s (the arc length) respect to the curve $l_r$ by the following formula Figure 4.

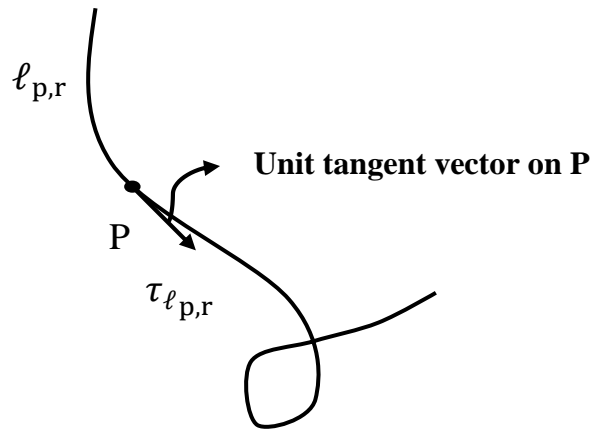

**Figure 4:** $\ell_{p,r}$, the reconstructed curve that acrosses from point P; $\tau_{p,r}$, the Unit tangent vector on P.

We define $\Gamma_{\ell_{p,r}}$ as the pure tension which acts through the curve $\ell_{p,r}$ on point P by the following formula:

$$\Gamma_{\ell_{p,r}} = \sum_{1 \leq i,j \leq 3} e_{ij,\ell_{p,r}} \cdot x_{i,r} \cdot x_{j,r}$$

$e_{ij,\ell_{p,r}}$'s are pure strain components at 3D space they provide a matrix "$\varepsilon = \left[e_{ij\ell_{p,r}}\right]_{3\times 3}$" that is named the strain matrix.

We can easily formulate the body force over the time at point P on $\ell_P$.

$$F_{\ell_{P,r}}(P,t) := \frac{\partial (\Gamma_{\ell_{p,r}} \cdot \tau_{\ell_{p,r}})}{\partial S_{\ell_{p,r}}}$$

Now let $\ell^* := \{\ell \text{ inside LV myocardium} \mid P \in \ell := \ell_{p,r} \text{ for some } r\}$, then total body force that is resulted in the pure strain and motion is formulated by:

$$F_P(P,t) := \sum_{\ell_{p,r} \in \ell^*} F_{\ell_{p,r}}(P,t)$$

$$\sum_{\ell_{p,r} \in \ell^*} F_{\ell_P}(P,t) = \sum_{\ell_{p,r} \in \ell^*} \frac{\partial \left(\Gamma_{\ell_{p,r}} \cdot \tau_{\ell_{p,r}}\right)}{\partial S_{\ell_{p,r}}} = \sum_{\ell_{p,r} \in \ell^*} \frac{\partial \left(\left(\sum_{1\leq i,j \leq 3} e_{ij,\ell_{p,r}} x_{i,r} x_{j,r}\right) \cdot \tau_{\ell_{p,r}}\right)}{\partial S_{\ell_{p,r}}}$$

$$F_P(P,t) = \sum_{\ell_{p,r} \in \ell^*} \frac{\partial \left(\left(\sum_{1\leq i,j \leq 3} e_{ij,p,r} x_i x_j\right) \cdot \tau_{\ell_{p,r}}\right)}{\partial S_{\ell_{p,r}}}$$

Body force at each point P in the LV myocardium are calculated by above formula based on Speckle Tracking software (ST software) to evaluate $e_{ij}$ and velocities and displacements. Let X be the set of all LV myocardial points P. we define the force map $\varphi$:

$$\varphi: X \to \mathbb{C}^3$$
$$P \mapsto (F_P, \Gamma_P, \tau_P)$$

$$\Gamma_P = \sum \Gamma_{\ell_{P,r}} = \sum_{\ell_{P,r}} \sum_{1\leq i,j\leq 3} e_{ij,\ell_{p,r}} x_{i,r} x_{j,r}$$

$\tau_P \to$ Normal vector on the surface of tangent vector at P.

$\varphi$ is a smooth and flat map; $\varphi^{-1}(z)$'s are myocardial fiber transactions for each $z \in \mathbb{C}^3$.

$$\varphi^{-1}(z) = \{P \in X \mid (F_P, \Gamma_P, \tau_P) = (z_1, z_2, z_3)\}$$

$$\Rightarrow X_{z,p} = \text{Spec}\left(\frac{\mathbb{C}[x,y,z]}{\langle F_P - z_1 \rangle \cdot \langle \Gamma_P - z_2 \rangle \cdot \langle \tau_P - z_3 \rangle}\right); \quad z \in \mathbb{C}^3, z_i \in \mathbb{C}.$$

$$X = \coprod_{\substack{z \in \mathbb{C}^3 \\ P \in X}} \text{Spec}\left(\frac{\mathbb{C}[x,y,z]}{\langle F_P - z_1 \rangle \cdot \langle \Gamma_P - z_2 \rangle \cdot \langle \tau_P - z_3 \rangle}\right)$$

Finally the left ventricle "X" is represented by a 3D scheme over complex numbers.

## Conclusion:

It was possible that the fiber movements to be discussed here were identical with the so-called "myocardial fiber transactions". It is hoped that some enquirer may succeed shortly in this study here, which is so important in connection with the theory of schemes.

## Acknowledgment:

The authors are really grateful to Professor David Mumford for his motivic question which stated "who one can explain schemes for biologists?"

## Disclosure:

There is no disclosure.